\documentclass[preprint,superscriptaddress,showpacs,nofootinbib,aps]{revtex4}
 \usepackage{graphicx}
 \usepackage{bm}
 
\begin{document}

 \title{Study of $B_{c}^{-}$ ${\to}$ $J/{\psi}{\pi}^{-}$, ${\eta}_{c}{\pi}^{-}$
        Decays with QCD Factorization}
 \author{Junfeng Sun}
 \affiliation{College of Physics and Information Engineering,
              Henan Normal University,
              Xinxiang 453007, China}
 \thanks{Mailing address}

 \author{Guifeng Xue}
 \affiliation{College of Physics and Information Engineering,
              Henan Normal University,
              Xinxiang 453007, China}

 \author{Yueling Yang}
 \affiliation{College of Physics and Information Engineering,
              Henan Normal University,
              Xinxiang 453007, China}

 \author{Gongru Lu}
 \affiliation{College of Physics and Information Engineering,
              Henan Normal University,
              Xinxiang 453007, China}

 \author{Dongsheng Du}
 \affiliation{Institute of High Energy Physics,
              Chinese Academy of Sciences,\\
              P.O.Box 918(4),
              Beijing 100049, China}

 \begin{abstract}
 The $B_{c}$ ${\to}$ $J/{\psi}{\pi}$, ${\eta}_{c}{\pi}$ decays are studied in
 the scheme of the QCD factorization approach. The branching ratios are calculated
 with the asymptotic distribution amplitude of the pion. The charm quark mass
 effect is considered. We find that the mass effect on the branching ratios is
 small.
 \end{abstract}
 \pacs{12.38.Bx  12.39.St  13.25.Hw}

 \maketitle

 \section{Introduction}
 \label{sec1}
 The $B_{c}$ meson is the ground state of the $b$-$c$ system. It is a quarkonium state
 consisting of heavy quarks with different flavours, and it lies below the threshold
 of the $BD$ mesons. It cannot annihilate into gluons, so it can only decay weakly.
 The $B_{c}$ system offers a new ideal place for studying the weak decay mechanism
 of heavy flavors and testing of the quark-flavour-mixing sector of the Standard Model.
 The discovery of the $B_{c}$ meson by the CDF \cite{9804014} has demonstrated the
 possibility for the experimental investigation of this system.

 $B_{c}$ mesons are too massive to produce at $e^{+}e^{-}$ colliders operating near
 ${\Upsilon}(4S)$, such as the ``$B$ factories'' at SLAC and KEK. They can be produced
 in significant numbers in hadron colliders.
 The Large Hadron Collider (LHC) is scheduled to run in 2008.
 Due to its high collision energy and high luminosity, the expected large production
 rates at the LHC will allow for precision measurements of $B_{c}$ properties.
 The $B$-physics, including the physics potential of the $B_{c}$ system, could be fully
 exploited at the LHC. Measurements of heavy quarkonia at the LHC will be valuable
 for a deep understanding of the production and decay mechanisms of the corresponding
 quarkonia involved.

 The $B_{c}$ meson has very rich weak decay channels which can be divided into
 three classes:
 (1) the $b$ quark decay ($b$ ${\to}$ $c$, $u$) with $c$ quark as a spectator,
 (2) the $c$ quark decay ($c$ ${\to}$ $s$, $d$) with $b$ quark as a spectator,
 and (3) the annihilation channel. Clearly, we can study the two heavy flavors
 $b$ and $c$ simultaneously with $B_{c}$ meson.
 For the case of Class (1), the precision determination of the Cabibbo-Kobayashi-Maskawa
 (CKM) \cite{ckm} matrix elements of ${\vert}V_{cb}{\vert}$ and ${\vert}V_{ub}{\vert}$
 are interesting, where determinations of ${\vert}V_{ub}{\vert}$ based on inclusive
 and exclusive channels have an uncertainty of 8\% and 17\% at present, respectively \cite{pdg2006}.
 For the case of Class (2), although the heavy $B_{s}$ (or $B_{u}$, $B_{d}$) in
 the final state has a considerable effect in reducing the phase space for $c$ quark
 decay, the CKM matrix elements ${\vert}V_{cb}{\vert}$ ${\ll}$ ${\vert}V_{cs}{\vert}$
 is in favor of the $c$ quark decay greatly. For the case of Class (3), the $B_{c}$
 weak annihilation decays are CKM enhanced by ${\vert}V_{cb}/V_{ub}{\vert}^{2}$ ${\sim}$ $10^{2}$
 as compared to $B_{u}$ annihilation decays.
 All the three classes listed above are interesting.

 The $B_{c}$ meson decay has been widely studied in the literature due to some of
 its outstanding features. The earlier nonleptonic decays of $B_{c}$ meson has been
 studied in \cite{prd39p1342,zpc51p549,zpc57p43,prd49p3399,prd56p4133,prd62p014019,npb585p353,
 PAN64p2027,JPG28p595,JPG30p1445,prd68p094020,plb555p189,prd73p054024,IJMPA21p777,07042493,0412158}.
 The theoretical status of the $B_{c}$ meson was reviewed in \cite{0412158}.
 In this paper, we will concentrate on the $B_{c}$ ${\to}$ $J/{\psi}{\pi}$, ${\eta}_{c}{\pi}$
 decays in Class (1) with QCD factorization approach. Let us outline a few reasons below.
 \begin{itemize}
 \item The LHCb detector has a high trigger performance, good resolution for interaction vertex,
       excellent particle identification for charged particles. So all final-state particles
       are detectable for the $B_{c}$ ${\to}$ $J/{\psi}{\pi}$, ${\eta}_{c}{\pi}$ decays.
       In practice, the most constructive information is to extract a signal of $J/{\psi}$ (or
       ${\eta}_{c}$) from the cascade decays \cite{0505076}. Especially, compared with the
       semi-leptonic decays where the neutrino momentum is not detected directly, the LHC can
       provide a clean signature for $B_{c}$ ${\to}$ $J/{\psi}{\pi}$, ${\eta}_{c}{\pi}$ due
       to the narrow-peak of $J/{\psi}$ (or ${\eta}_{c}$) and the good identification of
       the charged pion. It is estimated that one could expect around
       $5$ ${\times}$ $10^{10}$ $B_{c}$ events per year at LHC, so ATLAS would
       be able to record about 5600 events of $B_{c}$ ${\to}$ $J/{\psi}{\pi}$
       in one year \cite{0412158}. Better signal-to-background ratios
       and larger yields make $B_{c}$ ${\to}$ $J/{\psi}{\pi}$, ${\eta}_{c}{\pi}$
       decay modes to be the most prospective channels for measurements.
 \item The final states of $B_{c}$ ${\to}$ $J/{\psi}{\pi}$, ${\eta}_{c}{\pi}$ decays have
       large momentum compared with the Class (2) decays (e.g. $B_{c}$ ${\to}$ $B$ $+$ $X$
       decays), so the final state interactions can be neglected among the energetic particles.
       One can effectively factorize the hard and soft amplitudes. Further more, the behavior
       of interaction vertices modified by hard gluon corrections at large recoil can be
       calculated reliably with perturbation theory.
 \item For the exploration of $CP$ violation, the key processes are nonleptonic $B$ decays.
       The theoretical challenge is the calculation of the hadronic transition
       matrix elements ${\langle}f{\vert}\hat{Q}_{i}{\vert}B{\rangle}$, where $\hat{Q}_{i}$
       is the local four quark operator. Recently, several attractive methods have been
       proposed to study the ``nonfactorizable'' effects in hadronic matrix elements,
       such as the QCD factorization (QCDF) \cite{9905312}, perturbative QCD method
       \cite{9607214,9701233,0004004}, soft collinear effective theory
       \cite{prd63p114020,prd65p054022}. The nonleptonic two-body $B_{u}$, $B_{d}$, $B_{s}$
       decays have been studied in detail, for example, in Refs.
       \cite{9804363,9809364,9903453,0104090,0005006,0104110,0108141,prd68p054003,
       npb657p333,prd72p094026,plb620p143,npb774p64,0703162}.
       It is found that with appropriate parameters, most of the theoretical predictions
       are in agreement with the present experimental data. With the accumulation of data,
       the Standard Model can be tested in more detail. The potential $B_{c}$ decay modes
       permit us to over-constrain the angles and sides of the {\em unitarity triangle}
       of the CKM matrix. The $B_{c}$ ${\to}$ $J/{\psi}{\pi}$, ${\eta}_{c}{\pi}$ decays
       are $a_{1}$ dominant within the framework of Operator Product Expansion (OPE). They
       do not have pollution from penguins and annihilation diagrams. Moreover, the
       coefficient of $a_{1}$ can be well determined compared with other coefficients.
       So these decays could be used for determination of the CKM matrix element
       ${\vert}V_{cb}{\vert}$.
 \item In the limit of infinitely heavy quark masses, the HQET Lagrangians possess spin
       symmetry. The $B_{c}$ decays to $J/{\psi}$, ${\eta}_{c}$ involve the heavy spectator
       quark. The spectator enters the heavy hadron in the final states with no hard gluon
       rescattering, because spin symmetry works only at recoil momenta close to zero.
       The transition form factors should be nonperturbative hadronic quantities.
       It is expected that the quark-gluon sea is suppressed for $B_{c}$ mesons and that
       the calculation of factorization can be applied for non-leptonic $B_{c}$ decay.
       Thus the important parameter is the factor $a_{1}$ which depends on the normalization
       scale \cite{0412158}. The authors of
       \cite{prd39p1342,zpc51p549,zpc57p43,prd49p3399,prd56p4133,prd62p014019,npb585p353,
    PAN64p2027,JPG28p595,JPG30p1445,prd68p094020,plb555p189,prd73p054024,IJMPA21p777,07042493}
       have much concern for hadron transition form factors whose dependence on the momentum transfer
       is not significant within the physical phase space \cite{0412158}, rather than the
       factor $a_{1}$. Moreover, for the exploration of $CP$ violation in $B_{c}$ decays, it
       needs to know the strong phases in addition to weak phases. While only the real
       value of the factor $a_{1}$ at the fixed point is used in the previous work
       \cite{prd39p1342,zpc51p549,zpc57p43,prd49p3399,prd56p4133,prd62p014019,npb585p353,
    PAN64p2027,JPG28p595,JPG30p1445,prd68p094020,plb555p189,prd73p054024,IJMPA21p777,07042493}.
       In this paper, we will consider the radiative corrections to
       the hadron matrix element to compensate the renormalization scale dependence of Wilson
       coefficients. In addition, some information of strong phases can be obtained.
 \end{itemize}

 This paper is organized as follows:
 In section \ref{sec2}, we discuss the theoretical framework and compute the decay
 amplitudes for $B_{c}$ ${\to}$ $J/{\psi}{\pi}$, ${\eta}_{c}{\pi}$ with the master
 QCD formula.
 The section \ref{sec3} is devoted to the numerical results.
 Finally, we summarize in section \ref{sec4}.

 \section{Theoretical framework}
 \label{sec2}

 \subsection{The effective Hamiltonian}
 \label{sec21}
 Using the operator product expansion and renormalization group equation,
 the low energy effective Hamiltonian for $B_{c}$ ${\to}$ $J/{\psi}{\pi}$,
 ${\eta}_{c}{\pi}$ decays can be written as \cite{9512380}:
 \begin{equation}
 {\cal H}_{eff}\,=\,\frac{G_{F}}{\sqrt{2}} V_{cb}V_{ud}^{\ast} \Big\{
    C_{1}({\mu})Q_{1}+ C_{2}({\mu})Q_{2} \Big\} + \hbox{H.c.},
 \label{eq:Hamiltonian}
 \end{equation}
 where $V_{cb}V_{ud}^{\ast}$ is the CKM factors. The expressions of the
 local tree operators are
 \begin{equation}
 Q_{1}\,=\,(\bar{c}_{\alpha}b_{\alpha})_{V-A}(\bar{d}_{\beta}u_{\beta})_{V-A}~~~~~~~~~~
 Q_{2}\,=\,(\bar{c}_{\alpha}b_{\beta})_{V-A}(\bar{d}_{\beta}u_{\alpha})_{V-A}
 \label{eq:operator}
 \end{equation}
 Note that there is no local penguin operators, thus no interference between tree
 and penguin operators, and no $CP$ asymmetry for these decays. The coupling parameters
 $C_{i}({\mu})$ are Wilson coefficients which have been evaluated to
 the next-to-leading order (NLO). Their numerical values in the naive
 dimensional regularization scheme (NDR) are listed in Table \ref{tab1}.

 \subsection{Hadronic matrix elements within the QCDF framework}
 \label{sec22}
 For the weak decays of hadrons, the short-distance effects are well known
 and can be calculated with perturbation theory. However, the
 nonperturbative long-distance effects responsible for the hadronization from
 quarks to hadrons still remain obscure in several aspects. But to calculate
 the exclusive weak decays of the $B_{c}$ meson, one needs to evaluate the
 hadronic matrix elements, i.e., the weak current operator sandwiched between
 the initial state of the $B_{c}$ meson and the concerned hadronic final states,
 which is the most difficult theoretical work at present.
 Phenomenologically, these hadronic matrix elements are usually parameterized
 into the product of the decay constants and the transition form factors based
 on the naive factorization scheme (NF) \cite{bsw}. However, the main defect
 of the rough NF approach is that the hadronic matrix elements cannot compensate
 the renormalization scheme- and scale- dependence of Wilson coefficients.
 In that sense the NF's results are unphysical. Moreover, for the exploration of $CP$
 violation, information of the strong phases is needed, but the values of both
 the hadronic matrix elements and the Wilson coefficients are real within NF.
 These indicate that ``nonfactorizable'' contributions from high order corrections
 to the hadronic matrix elements must be taken into account.

 A few years ago, Beneke, Buchalla, Neubert, and Sachrajda suggested a QCDF
 formula to compute the hadronic matrix elements
 ${\langle}M_{1}M_{2}{\vert}O_{i}{\vert}B{\rangle}$
 in the heavy quark limit, combining the hard scattering approach with
 power counting in $1/m_{b}$ \cite{9905312}. At leading order in the
 heavy quark expansion, the hadronic matrix elements can be factorized
 into ``non-factorizable'' corrections dominated by hard gluon exchange
 and universal non-perturbative part parameterized by the physical form
 factors and meson's light cone distribution amplitudes. This promising
 method has been applied to exclusive two-body non-leptonic $B_{u}$,
 $B_{d}$, $B_{s}$ decays \cite{0108141,prd68p054003,npb657p333}.
 It is found that with appropriate parameters, most of the QCDF's
 predictions are in agreement with the present experimental data.
 In this paper, we would like to apply the QCDF approach to the case of
 $B_{c}$ ${\to}$ $J/{\psi}{\pi}$, ${\eta}_{c}{\pi}$ decays.

 In the heavy quark limit $m_{b}$ $\gg$ ${\Lambda}_{QCD}$, up to power corrections
 of order of ${\Lambda}_{QCD}/m_{b}$, the master QCDF formula for $B_{c}$ ${\to}$
 ${\eta}_{c}{\pi}$ (or $J/{\psi}{\pi}$) can be written as \cite{9905312}
 \begin{equation}
 {\langle}{\eta}_{c}{\pi}{\vert}O_{i}{\vert}B_{c}{\rangle}\ =\
  F^{B_{c}{\to}{\eta}_{c}} {\int}_{0}^{1}dz \ H(z) {\Phi}_{\pi}(z)
 \label{eq:qcdf}
 \end{equation}
 Through the QCDF formula Eq.(\ref{eq:qcdf}), the hadronic matrix elements can be separated
 into short-distance part [hard-scattering kernels $H(z)$] and long-distance part.
 At leading order of ${\alpha}_{s}$, $H(z)$ $=$ $1$. There is no long-distance interaction
 between pion and $B_{c}$-${\eta}_{c}$ (or $B_{c}$-$J/{\psi}$) system, so the NF's picture is
 recovered. Nonperturbative effects are either suppressed by $1/m_{b}$ or parameterized in
 terms of the transition form factors $F^{B_{c}{\to}{\eta}_{c}}$ (or $F^{B_{c}{\to}J/{\psi}}$)
 and the meson's light-cone distribution amplitudes ${\Phi}(z)$.

 \subsection{Discussions on soft and collinear divergence cancellation at one-loop order}
 \label{sec23}
\begin{figure}[ht]
 \includegraphics[200,670][400,750]{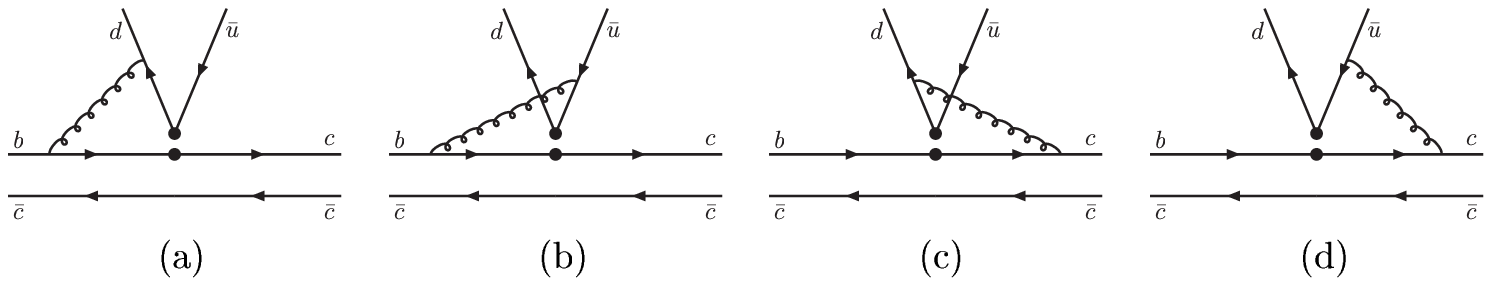}
 \caption{Vertex corrections to hard-scattering kernel for $b$ ${\to}$ $c$
 decay at the order of ${\alpha}_{s}$. The upward lines represent the valence
 quarks of the emitted ${\pi}$ meson.}
 \label{fig1}
 \end{figure}

 When we calculate the vertex corrections in the leading power of ${\Lambda}_{QCD}/m_{b}$,
 not only ultraviolet divergence emerges but also infrared divergence, as shown
 in \cite{0102077}. The authors of \cite{9905312} gave an explicit cancellation of soft
 and collinear divergences in vertex corrections for $B$ ${\to}$ $D{\pi}$ decays. They
 demonstrated that the soft divergences cancel out in Fig.\ref{fig1}(a) and
 Fig.\ref{fig1}(b), and in Fig.\ref{fig1}(c) and Fig.\ref{fig1}(d), respectively;
 the collinear divergences cancel out in Fig.\ref{fig1}(a) and Fig.\ref{fig1}(c),
 and in Fig.\ref{fig1}(b) and Fig.\ref{fig1}(d), respectively. For $B_{c}$ ${\to}$
 $J/{\psi}{\pi}$, ${\eta}_{c}{\pi}$ decays, the soft and collinear divergence cancellations
 are the same as that for $B$ ${\to}$ $D{\pi}$ decays in \cite{9905312}, because they are
 all $b$ ${\to}$ $c$ transition. Here, we will consider the non-zero mass of $c$ quark,
 and give an explicit calculation of the Feynman diagrams Fig.\ref{fig1}(a)-(d) to
 show the cancellation of the infrared divergence, as shown in \cite{0102077}.

 The two valence quarks in pion have transverse momentum as well as
 longitudinal momentum. The quark and antiquark momenta in pion meson is
 \begin{equation}
 k_{q}\ =\ zq+\vec{k}_{\perp},~~~~~~~~~~k_{\bar{q}}=\bar{z}q-\vec{k}_{\perp}
 \end{equation}
 where $q$ ${\sim}$ ${\cal O}(m_{b})$ is the momentum of the emitted ${\pi}$
 meson; $z$ and $\bar{z}$ ${\equiv}$ $1$ $-$ $z$ are the longitudinal momentum
 fraction of quark and anti-quark, respectively; $\vec{k}_{\perp}$ ${\sim}$
 ${\cal O}({\Lambda}_{QCD})$ is the transverse momentum.
 But in $B_{c}$ ${\to}$ $J/{\psi}{\pi}$, ${\eta}_{c}{\pi}$ decays, the pion is
 energetic. So the dominant contribution to the pion wave function comes from
 configurations where two valence quarks are hard. The contribution from soft
 regions at the endpoint is believed to be suppressed in the heavy-quark limit.
 For our case, the quark is on-shell, and the effects proportional to
 $\vec{k}_{\perp}$ is power suppressed.

 Now we shall take into account the ${\alpha}_{s}$ corrections to the hadronic
 matrix element. After a straightforward calculation in the NDR scheme as in
 \cite{0102077}, we obtain
 \begin{eqnarray}
 \lefteqn{ {\langle}Q_{2}{\rangle}_{(a)} =
     \frac{{\alpha}_{s}}{4{\pi}}\frac{C_{F}}{N}
    {\langle}{\pi}^{-}{\vert}(\bar{d}u)_{V-A}{\vert}0{\rangle}
    {\langle}J/{\psi}{\vert}(\bar{c}b)_{V-A}{\vert}B_{c}^{-}{\rangle} }
  \nonumber \\ &{\times}& \Big\{
    -\Big( \frac{m_{b}^{2}}{{\mu}^{2}} \Big)^{\epsilon}
     \frac{{\Gamma}(1-{\epsilon})}{\big(4{\pi}\big)^{\epsilon}}
     \Big[ \frac{1}{{\epsilon}^{2}}+\frac{2}{\epsilon}\big({\ln}t_{a}-1\big)
     +{\ln}^{2}t_{a}+\frac{2{\ln}t_{a}}{1-t_{a}}-4{\ln}t_{a}
     -2{\rm Li}_{2}\Big(\frac{t_{a}-1}{t_{a}}\Big)+5
  \nonumber \\ & & ~~~~~~~~~~~~~~~~~~~~~~~~~~~~~~+
      r_{c}\Big( \frac{t_{a}{\ln}t_{a}}{(1-t_{a})^2}
     +\frac{1}{1-t_{a}} \Big) \Big]
  \nonumber \\ & & ~~
     +\Big( \frac{{\mu}^{2}}{m_{b}^{2}} \Big)^{{\varepsilon}}
     \big( 4{\pi} \big)^{{\varepsilon}}
    {\Gamma}(1+{\varepsilon})
     \Big[ \frac{1}{{\varepsilon}}+\frac{t_{a}{\ln}t_{a}}{1-t_{a}}+1 \Big] \Big\}
 \label{eq:figa} \\
 \lefteqn{ {\langle}Q_{2}{\rangle}_{(b)} =
     \frac{{\alpha}_{s}}{4{\pi}}\frac{C_{F}}{N}
    {\langle}{\pi}^{-}{\vert}(\bar{d}u)_{V-A}{\vert}0{\rangle}
    {\langle}J/{\psi}{\vert}(\bar{c}b)_{V-A}{\vert}B_{c}^{-}{\rangle} }
  \nonumber \\ &{\times}& \Big\{
    +\Big( \frac{m_{b}^{2}}{{\mu}^{2}} \Big)^{\epsilon}
     \frac{{\Gamma}(1-{\epsilon})}{\big(4{\pi}\big)^{\epsilon}}
     \Big[ \frac{1}{{\epsilon}^{2}}+\frac{2}{\epsilon}\big({\ln}t_{b}-1\big)
     +{\ln}^{2}t_{b}+\frac{2{\ln}t_{b}}{1-t_{b}}-4{\ln}t_{b}
     -2{\rm Li}_{2}\Big(\frac{t_{b}-1}{t_{b}}\Big)+6 \Big]
  \nonumber \\ & &~~
    -\Big( \frac{{\mu}^{2}}{m_{b}^{2}} \Big)^{{\varepsilon}}
     \big( 4{\pi} \big)^{{\varepsilon}}
    {\Gamma}(1+{\varepsilon})
     \Big[ \frac{4}{{\varepsilon}}+4\frac{t_{b}{\ln}t_{b}}{1-t_{b}}+11 \Big] \Big\}
 \label{eq:figb} \\
 \lefteqn{ {\langle}Q_{2}{\rangle}_{(c)} =
     \frac{{\alpha}_{s}}{4{\pi}}\frac{C_{F}}{N}
    {\langle}{\pi}^{-}{\vert}(\bar{d}u)_{V-A}{\vert}0{\rangle}
    {\langle}J/{\psi}{\vert}(\bar{c}b)_{V-A}{\vert}B_{c}^{-}{\rangle} }
  \nonumber \\ &{\times}& \Big\{
    +\Big( \frac{m_{c}^{2}}{{\mu}^{2}} \Big)^{\epsilon}
     \frac{{\Gamma}(1-{\epsilon})}{\big(4{\pi}\big)^{\epsilon}}
     \Big[ \frac{1}{{\epsilon}^{2}}+\frac{2}{\epsilon}\big({\ln}t_{c}-1\big)
     +{\ln}^{2}t_{c}+\frac{2{\ln}t_{c}}{1-t_{c}}-4{\ln}t_{c}
     -2{\rm Li}_{2}\Big(\frac{t_{c}-1}{t_{c}}\Big)+6 \Big]
  \nonumber \\ & &~~
     -\Big( \frac{{\mu}^{2}}{m_{c}^{2}} \Big)^{{\varepsilon}}
     \big( 4{\pi} \big)^{{\varepsilon}}
    {\Gamma}(1+{\varepsilon})
     \Big[ \frac{4}{{\varepsilon}}+4\frac{t_{c}{\ln}t_{c}}{1-t_{c}}+11 \Big] \Big\}
 \label{eq:figc} \\
 \lefteqn{ {\langle}Q_{2}{\rangle}_{(d)} =
     \frac{{\alpha}_{s}}{4{\pi}}\frac{C_{F}}{N}
    {\langle}{\pi}^{-}{\vert}(\bar{d}u)_{V-A}{\vert}0{\rangle}
    {\langle}J/{\psi}{\vert}(\bar{c}b)_{V-A}{\vert}B_{c}^{-}{\rangle} }
  \nonumber \\ &{\times}& \Big\{
    -\Big( \frac{m_{c}^{2}}{{\mu}^{2}} \Big)^{\epsilon}
     \frac{{\Gamma}(1-{\epsilon})}{\big(4{\pi}\big)^{\epsilon}}
     \Big[ \frac{1}{{\epsilon}^{2}}+\frac{2}{\epsilon}\big({\ln}t_{d}-1\big)
     +{\ln}^{2}t_{d}+\frac{2{\ln}t_{d}}{1-t_{d}}-4{\ln}t_{d}
     -2{\rm Li}_{2}\Big(\frac{t_{d}-1}{t_{d}}\Big)+5
  \nonumber \\ & & ~~~~~~~~~~~~~~~~~~~~~~~~~~~~~~+
      \frac{1}{r_{c}}\Big( \frac{t_{d}{\ln}t_{d}}{(1-t_{d})^2}
     +\frac{1}{1-t_{d}} \Big) \Big]
  \nonumber \\ & & ~~
     +\Big( \frac{{\mu}^{2}}{m_{c}^{2}} \Big)^{{\varepsilon}}
     \big( 4{\pi} \big)^{{\varepsilon}}
    {\Gamma}(1+{\varepsilon})
     \Big[ \frac{1}{{\varepsilon}}+\frac{t_{d}{\ln}t_{d}}{1-t_{d}}+1 \Big] \Big\}
 \label{eq:figd}
 \end{eqnarray}
 where ${\varepsilon}$ ${\to}$ $0^{+}$ for ultraviolet divergences,
 ${\epsilon}$ ${\to}$ $0^{-}$ for infrared divergences,
 $r_{c}$ $=$ $m_{c}/m_{b}$, and
 \begin{eqnarray}
 t_{a} &=& z(1-r_{c}^{2})  \\
 t_{b} &=& \bar{z}(1-r_{c}^{2})  \\
 t_{c} &=& z(1-1/r_{c}^{2})\ =\ -t_{a}/r_{c}^{2} \\
 t_{d} &=& \bar{z}(1-1/r_{c}^{2})\ =\ -t_{b}/r_{c}^{2}
 \end{eqnarray}

 In Eq.(\ref{eq:figc}), the term of infrared divergences can be written as
 \begin{eqnarray}
 & & \Big( \frac{m_{c}^{2}}{{\mu}^{2}} \Big)^{\epsilon}
     \Big[ \frac{1}{{\epsilon}^{2}}+\frac{2}{\epsilon}\big({\ln}t_{c}-1\big)
         +{\cdots} \Big]
 \nonumber \\ &=& \Big( \frac{m_{b}^{2}}{{\mu}^{2}} \Big)^{\epsilon}
     \Big( \frac{m_{c}^{2}}{m_{b}^{2}} \Big)^{\epsilon}
     \Big[ \frac{1}{{\epsilon}^{2}}+\frac{2}{\epsilon}
     \big({\ln}\frac{-t_{a}}{r_{c}^{2}}-1\big)+{\cdots} \Big]
 \nonumber \\ &=& \Big( \frac{m_{b}^{2}}{{\mu}^{2}} \Big)^{\epsilon}
     \Big[ \frac{1}{{\epsilon}^{2}}+\frac{2}{\epsilon}\big({\ln}t_{a}-1\big)
         -\frac{1}{\epsilon}\big({\ln}r^{2}+i\,2{\pi} \big)
         +{\cdots} \Big]
 \label{eq:figc-infrared}
 \end{eqnarray}

 We can treat the infrared divergences in Eq.(\ref{eq:figd}) in the same way.
 \begin{eqnarray}
 & & \Big( \frac{m_{c}^{2}}{{\mu}^{2}} \Big)^{\epsilon}
     \Big[ \frac{1}{{\epsilon}^{2}}+\frac{2}{\epsilon}\big({\ln}t_{d}-1\big)
         +{\cdots} \Big]
 \nonumber \\ &=& \Big( \frac{m_{b}^{2}}{{\mu}^{2}} \Big)^{\epsilon}
     \Big( \frac{m_{c}^{2}}{m_{b}^{2}} \Big)^{\epsilon}
     \Big[ \frac{1}{{\epsilon}^{2}}+\frac{2}{\epsilon}
     \big({\ln}\frac{-t_{b}}{r_{c}^{2}}-1\big)+{\cdots} \Big]
 \nonumber \\ &=& \Big( \frac{m_{b}^{2}}{{\mu}^{2}} \Big)^{\epsilon}
     \Big[ \frac{1}{{\epsilon}^{2}}+\frac{2}{\epsilon}\big({\ln}t_{b}-1\big)
         -\frac{1}{\epsilon}\big({\ln}r^{2}+i\,2{\pi} \big)
         +{\cdots} \Big]
 \label{eq:figd-infrared}
 \end{eqnarray}
 So it is easy to see that the infrared divergences are cancelled
 in the summation of all diagrams in Fig.\ref{fig1} before we integrate
 over the momentum fraction variable $z$. That is to say that the decay
 amplitude is infrared finite in the soft region when $\vec{k}_{\perp}$
 is neglected. This can be understood as a technical manifestation of
 colour transparency.

 Within the modified minimal subtraction ($\overline{\rm MS}$) scheme, we get
 \begin{eqnarray}
  & & {\langle}Q_{2}{\rangle}_{(a)+(b)+(c)+(d)}
  \nonumber \\ &=&
     \frac{{\alpha}_{s}}{4{\pi}}\frac{C_{F}}{N}
    {\langle}{\pi}^{-}{\vert}(\bar{d}u)_{V-A}{\vert}0{\rangle}
    {\langle}J/{\psi}{\vert}(\bar{c}b)_{V-A}{\vert}B_{c}^{-}{\rangle}
  \nonumber \\ &{\times}& \Big\{
     \Big[ \frac{t_{a}{\ln}t_{a}}{1-t_{a}}-\frac{4t_{b}{\ln}t_{b}}{1-t_{b}}
     +3{\ln}\Big( \frac{m_{b}^{2}}{{\mu}^{2}} \Big)-9 \Big]
     -r_{c} \Big[ \frac{t_{a}{\ln}t_{a}}{(1-t_{a})^2}+\frac{1}{1-t_{a}} \Big]
  \nonumber \\ & & +
     \Big[ \frac{t_{d}{\ln}t_{d}}{1-t_{d}}-\frac{4t_{c}{\ln}t_{c}}{1-t_{c}}
     +3{\ln}\Big( \frac{m_{c}^{2}}{{\mu}^{2}} \Big)-9 \Big]
     -\frac{1}{r_{c}} \Big[ \frac{t_{d}{\ln}t_{d}}{(1-t_{d})^2}
     +\frac{1}{1-t_{d}} \Big]
  \nonumber \\ & &
     +2({\ln}t_{a}-{\ln}t_{b}){\ln}r_{c}^{2}-f(t_{a})+f(t_{b})+f(t_{c})-f(t_{d}) \Big\}
 \label{eq:fig-sum}
 \end{eqnarray}
 where
 \begin{equation}
 f(t)\ =\ {\ln}^{2}t+\frac{2t{\ln}t}{1-t}-2{\rm Li}_{2}\Big(\frac{t-1}{t}\Big)
 \label{eq:f}
 \end{equation}

 From above we found that the expressions of Eq.(\ref{eq:fig-sum}) and
 Eq.(\ref{eq:f}) are consistent with the result [see Eq.(80)---Eq.(85)]
 in the previous works \cite{9905312}.

 With Eq.(\ref{eq:fig-sum}) and Eq.(\ref{eq:f}), we can compute the
 ``non-factorizable'' contribution of the one-gluon exchange vertex
 correction no matter the distribution amplitude ${\phi}(z)$
 is symmetric or not. This is very important in principle.
 For instance, when kaon is ejected from $b$ quark decay,
 the contribution from the asymmetric part of kaon distribution
 amplitude must be taken into account, although the contributions
 from the asymmetric part might be very small numerically.

 From the above, we can see that the ``nonfactorizable'' interactions
 connecting the $B_{c}$ ${\to}$ $J/{\psi}$ (or ${\eta}_{c}$) transition and
 the emitted pion are dominated by hard gluon exchange. So we may perform
 the $\vec{k}_{\perp}$ integration over the pion wave function.
 We define \cite{9905312}
 \begin{eqnarray}
 {\langle}{\pi}(q){\vert}\bar{d}_{\alpha}(x)u_{\beta}(0){\vert}0{\rangle}
 &=& {\int}{\bf d}z\,\frac{{\bf d}^{2}\vec{k}_{\perp}}{16{\pi}^{3}}
  \frac{e^{ik_{d}{\cdot}x}}{\sqrt{2N_{c}}}{\Psi}_{\pi}(z,\vec{k}_{\perp})
  \Big[ {\gamma}_{5}\not{q} \Big]_{{\alpha}{\beta}}
  \nonumber \\ &=& -\frac{if_{\pi}}{4N_{c}}{\int}{\bf d}z\,{\phi}_{\pi}(z)
  \Big[ {\gamma}_{5}\not{q} \Big]_{{\alpha}{\beta}}
 \label{eq:pion-wave-function}
 \end{eqnarray}

 Assuming that the distribution amplitude ${\phi}_{\pi}(z)$ is symmetric,
 i.e. ${\phi}_{\pi}(z)$ $=$ ${\phi}_{\pi}(\bar{z})$, then Eq.(\ref{eq:fig-sum})
 can be simplified as follows:
 \begin{eqnarray}
  & & {\langle}Q_{2}{\rangle}_{(a)+(b)+(c)+(d)}
  \nonumber \\ &=&
     \frac{{\alpha}_{s}}{4{\pi}}\frac{C_{F}}{N}
    {\langle}{\pi}^{-}{\vert}(\bar{d}u)_{V-A}{\vert}0{\rangle}
    {\langle}J/{\psi}{\vert}(\bar{c}b)_{V-A}{\vert}B_{c}^{-}{\rangle}
  \nonumber \\ &{\times}& \Big\{ 3\Big[
     {\ln}\Big( \frac{m_{b}^{2}}{{\mu}^{2}} \Big)
    +{\ln}\Big( \frac{m_{c}^{2}}{{\mu}^{2}} \Big)
    -\frac{t_{a}{\ln}t_{a}}{1-t_{a}}
    -\frac{t_{d}{\ln}t_{d}}{1-t_{d}}
    -6 \Big]
  \nonumber \\ & &\;
     -\Big[r_{c}\Big( \frac{t_{a}{\ln}t_{a}}{(1-t_{a})^2}
     +\frac{1}{1-t_{a}} \Big)+\frac{1}{r_{c}}
      \Big( \frac{t_{d}{\ln}t_{d}}{(1-t_{d})^2}
     +\frac{1}{1-t_{d}} \Big) \Big] \Big\}
 \label{eq:fig-sum-symmetric}
 \end{eqnarray}

 Now we present the results of the coefficients $a_{1}$ for the case
 of $B_{c}$ ${\to}$ $J/{\psi}{\pi}$, ${\eta}_{c}{\pi}$ decays.
 Similar results hold for all Class (1) ($a_{1}$-dominant) $b$
 ${\to}$ $cqq^{\prime}$ (where $q$ and $q^{\prime}$ are light quarks)
 transition processes.
 \begin{equation}
  a_{1}\ =\ C^{^{\rm (NLO)}}_{1}+\frac{C^{^{\rm (NLO)}}_{2}}{N_{c}}
         + \frac{{\alpha}_{s}}{4{\pi}}\frac{C_{F}}{N_{c}}C^{^{\rm (LO)}}_{2}V
 \label{eq:a1}
 \end{equation}

 The vertex corrections $V$are given as follows
 [note: here ${\phi}(z)$ $=$ ${\phi}(\bar{z})$ is assumed]:
 \begin{itemize}
 \item for the $m_{c}$ ${\neq}$ $0$ case
 \begin{equation}
  V\ =\ 3{\ln}\Big( \frac{m_{b}^{2}}{{\mu}^{2}} \Big)
      + 3{\ln}\Big( \frac{m_{c}^{2}}{{\mu}^{2}} \Big)
      -18-{\int}{\bf d}z\, {\phi}(z)H_{1}(z)
  \label{eq:vmc}
  \end{equation}
  where
  \begin{equation}
  H_{1}(z)\ =\ \frac{3t_{a}{\ln}t_{a}}{1-t_{a}}
            +  \frac{3t_{d}{\ln}t_{d}}{1-t_{d}}
            +  r_{c} \Big( \frac{t_{a}{\ln}t_{a}}{(1-t_{a})^2}
                         + \frac{1}{1-t_{a}} \Big)
            +  \frac{1}{r_{c}} \Big( \frac{t_{d}{\ln}t_{d}}{(1-t_{d})^2}
                                   + \frac{1}{1-t_{d}} \Big)
 \label{eq:h}
 \end{equation}
 \item for the $m_{c}$ $=$ $0$ case \cite{0104110,0102077}
 \begin{equation}
  V\ =\ 3 \Big\{ 2{\ln}\Big( \frac{m_{b}^{2}}{{\mu}^{2}}\Big)-6-i{\pi}
       +{\int}{\bf d}z\, {\phi}(z) \frac{\bar{z}-z}{1-z}{\ln}z \Big\}
 \label{eq:vm0}
 \end{equation}
 \end{itemize}

 \subsection{The renormalization scale dependence}
 \label{sec24}
 From the expressions of the coefficient $a_{1}$ in Eq.(\ref{eq:a1}),
 apparently, the renormalization scale dependence of the hadronic matrix
 element is recovered, i.e. ${\langle}Q_{2}({\mu}){\rangle}$ $=$
 $g({\mu}){\langle}Q_{2}{\rangle}$. We expect that the recovered
 scale dependence from $g({\mu})$ can cancel the scale dependence
 of the Wilson coefficients $C_{i}({\mu})$, at least in part.

\begin{figure}[ht]
 \includegraphics[250,380][350,550]{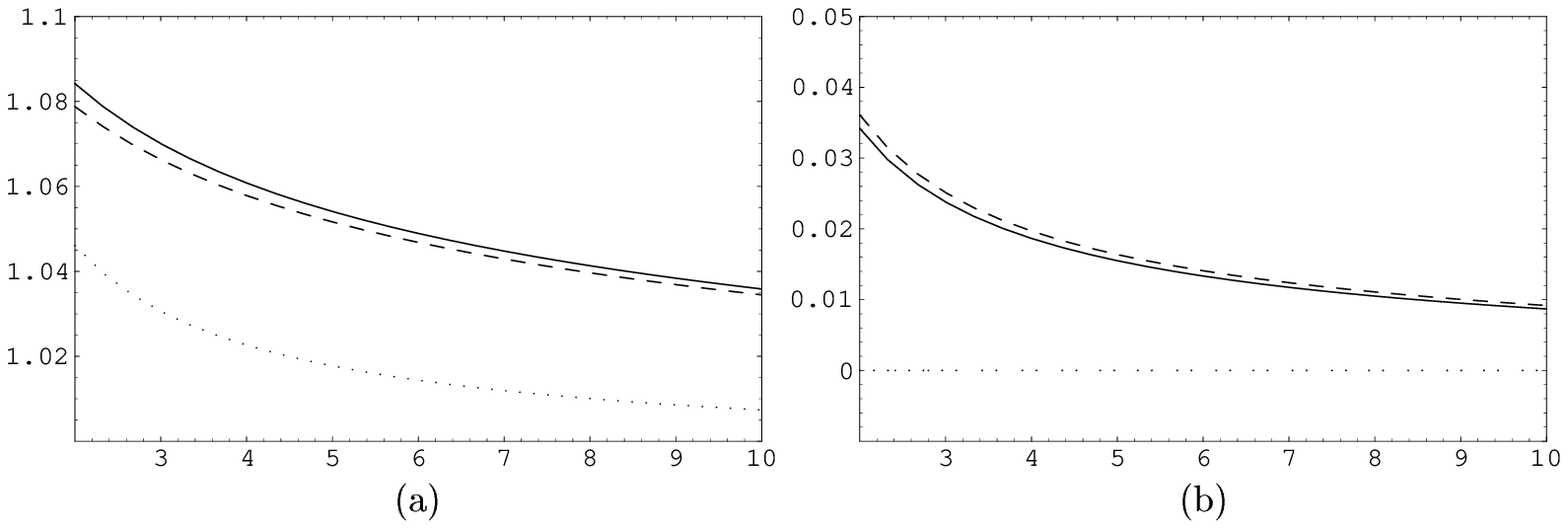}
 \caption{Dependence of the coefficient $a_{1}$ [vertical axes, ${\rm Re}(a_{1})$ in (a) and
  ${\rm Im}(a_{1})$ in (b)] on the renormalization scale ${\mu}$ [horizontal axes, in units
  of GeV], with asymptotic light-cone distribution amplitudes ${\phi}_{\pi}(x)$ $=$ $6x\bar{x}$.
  The solid and dashed lines denote $a_{1}$ for the case of $m_{c}$ $=$ $1.25$ GeV and
  $m_{c}$ $=$ $0$ within QCDF framework, respectively; the dotted line denotes $a_{1}$
  within NF framework.}
 \label{fig2}
 \end{figure}

 With the renormalization group equations for the Wilson coefficients
 at leading order (LO) approximation \cite{9512380}
 \begin{equation}
 \frac{ {\bf d} }{ {\bf d}{\ln}{\mu} }
 \left( \begin{array}{c} C_{1}({\mu}) \\ C_{2}({\mu}) \end{array} \right)\ =\
 \frac{{\alpha}_{s}}{4{\pi}}{\gamma}_{s}^{(0)T}
 \left( \begin{array}{c} C_{1}({\mu}) \\ C_{2}({\mu}) \end{array} \right)
 \label{eq:renormalization-group-equations}
 \end{equation}
 where the matrix of the LO anomalous dimensions ${\gamma}_{s}^{(0)}$ is \cite{9512380}
 \begin{equation}
 {\gamma}^{(0)}_{s}\ =\ \left( \begin{array}{cc}
  -6/N_{c} & 6 \\ 6 & -6/N_{c} \end{array} \right)
 \label{eq:anomalous-dimensions}
 \end{equation}
 It is easy to find that when the contributions from higher
 order of ${\alpha}_{s}$ are neglected, we have
 \begin{equation}
  \frac{{\bf d}}{{\bf d}{\ln}{\mu}}a_{1}\ =\
  \frac{{\bf d}}{{\bf d}{\ln}{\mu}}\Big(C_{1}({\mu})+\frac{C_{2}({\mu})}{N_{c}}\Big)
 +\frac{{\alpha}_{s}}{4{\pi}}\frac{C_{2}({\mu})}{N_{c}}C_{F}
  \frac{{\bf d}}{{\bf d}{\ln}{\mu}}V\ =\ 0
 \label{eq:scale-dependence}
 \end{equation}
 Eq.(\ref{eq:scale-dependence}) just show that the coefficient $a_{1}$ is no longer
 scale-dependent at the order of ${\alpha}_{s}$. Of course, $a_{1}$ is still scale-dependent
 beyond the order of ${\alpha}_{s}$ corrections. This can be seen from both Fig.\ref{fig2}
 and the numbers in Table \ref{tab1}. In principle, we have to include contributions of all
 higher order of ${\alpha}_{s}$ corrections to cancel the remaining scale dependence of $a_{1}$.

 \subsection{Gauge dependence}
 \label{sec25}
 To demonstrate the gauge dependence, firstly we would like to write down
 the one-gluon exchange contribution to $B_{c}$ ${\to}$ $J/{\psi}{\pi}$
 matrix element of the operator $Q_{2}$. Following the expression of
 Eq.(55)---Eq.(57) in \cite{9905312}, we have
 \begin{eqnarray}
 {\langle}Q_{2}{\rangle}_{{\rm Fig.}\ref{fig1}}
 = -i\frac{g_{s}^{2}C_{F}}{2} {\int}\frac{{\bf d}^{4}k}{(2{\pi})^4}
 {\langle}J/{\psi}{\vert}\bar{c}A^{\alpha}_{1}(l)b{\vert}B_{c}{\rangle}
  \frac{g_{{\alpha}{\beta}}}{l^{2}} {\int}{\bf d}z
  \frac{{\bf d}^{2}\vec{k}_{\perp}}{16{\pi}^{3}}
  \frac{{\Psi}_{\pi}(z,\vec{k}_{\perp})}{\sqrt{2N_{c}}}
 {\bf Tr}\Big[ {\gamma}_{5}\!\!\not{q}A^{\beta}_{2}(l,k_{d},k_{\bar{u}}) \Big]
 \label{eq:fig1}
 \end{eqnarray}
 where $l$ is the momentum of the internal gluon propagator, and
 \begin{eqnarray}
 && A^{\alpha}_{1}(l)\ =\ {\gamma}^{\alpha}
    \frac{\not{p}_{c}-\not{l}+m_{c}}{(p_{c}-l)^{2}-m_{c}^{2}}
   {\Gamma}+{\Gamma}
    \frac{\not{p}_{b}+\not{l}+m_{b}}{(p_{b}+l)^{2}-m_{b}^{2}}
   {\gamma}^{\alpha}
 \label{eq:fig1-a1} \\
 && A^{\beta}_{2}(l,k_{d},k_{\bar{u}})\ =\ {\Gamma}
    \frac{\not{k}_{\bar{u}}+\not{l}}{(k_{\bar{u}}+l)^{2}}
   {\gamma}^{\beta}-{\gamma}^{\beta}
    \frac{\not{k}_{d}+\not{l}}{(k_{d}+l)^{2}}
   {\Gamma}
 \label{eq:fig1-a2}
 \end{eqnarray}
 Here ${\Gamma}$ $=$ ${\gamma}_{\mu}(1-{\gamma}_{5})$; $p_{b}$ and $p_{c}$
 are the momentum of the $b$ quark and the $c$ quark, respectively.
 In fact, the terms in Eq.(\ref{eq:fig1}) are calculated using Feynman gauge.
 For an arbitrary covariant gauge, the internal gluon propagator should be
 written as \cite{QCD-field}
 \begin{equation}
 \frac{-i}{l^{2}}\Big(g_{{\alpha}{\beta}}+{\eta}\frac{l_{\alpha}l_{\beta}}{l^{2}}\Big)
 \label{eq:propagator}
 \end{equation}
 where ${\eta}$ $=$ $0$ is the Feynman gauge, ${\eta}$ $=$ $-1$ is the Landau gauge.

 Assuming that the external states are all physical and can be approximated
 as on-shell quarks in the leading order of ${\Lambda}_{QCD}/m_{b}$,
 it is  easy to verify that for arbitrary covariant gauge, the terms
 proportional to ${\eta}$ in Eq.(\ref{eq:fig1}) have no contribution,
 i.e. the expression of Eq.(\ref{eq:fig1}) is gauge independent.
 For instance,
 \begin{eqnarray}
 l_{\alpha}l_{\beta}A^{\alpha}_{1}(l) &=&
 l_{\beta}\not{l} \frac{\not{p}_{c}-\not{l}+m_{c}}{(p_{c}-l)^{2}-m_{c}^{2}}
 {\Gamma}+{\Gamma} \frac{\not{p}_{b}+\not{l}+m_{b}}{(p_{b}+l)^{2}-m_{b}^{2}}
  \not{l}l_{\beta} \nonumber \\ &=&
 -l_{\beta} \frac{2p_{c}{\cdot}l-l^{2}}{2p_{c}{\cdot}l-l^{2}}{\Gamma}
 +{\Gamma} \frac{2p_{b}{\cdot}l+l^{2}}{2p_{b}{\cdot}l+l^{2}} l_{\beta}\ =\ 0
 \label{eq:gauge-a1}
 \end{eqnarray}
 Hence the gauge dependence cancels out when adding the two terms in $A_{1}$.
 We can treat the $A_{2}$ in the similar way and get the result,
 $l_{\alpha}l_{\beta}A^{\beta}_{2}$ $=$ $0$. More precisely, we find that
 the expression of Eq.(\ref{eq:fig1}) is gauge independent.

 \section{Numerical results and discussions}
 \label{sec3}
 The expressions of decay amplitudes for $B_{c}$ ${\to}$ $J/{\psi}{\pi}$,
 ${\eta}_{c}{\pi}$ within the QCDF framework can be written as
 \begin{eqnarray}
 {\cal A}(B_{c}^{-}{\to}J/{\psi}{\pi}^{-})&=&\sqrt{2}\,G_{F}
   V_{cb}V_{ud}^{\ast} f_{\pi}A_{0}^{B_{c}{\to}J/{\psi}}
   m_{_{J/{\psi}}}({\epsilon}_{_{J/{\psi}}}{\cdot}p_{_{\pi}})a_{1}
 \label{eq:amp-jpsi} \\
 {\cal A}(B^{-}_{c}{\to}{\eta}_{c}{\pi}^{-})&=&-i\frac{G_{F}}{\sqrt{2}}
   V_{cb}V_{ud}^{\ast} f_{\pi}F_{0}^{B_{c}{\to}{\eta}_{c}}
   (m_{_{B_{c}}}^{2}-m_{_{{\eta}_{c}}}^{2})a_{1}
 \label{eq:amp-etac}
 \end{eqnarray}

 The branching ratios in $B_{c}$ meson rest frame can be written as:
 \begin{equation}
  BR(B_{c}{\to}X_{c\bar{c}}{\pi})= \frac{{\tau}_{B_{c}}}{8{\pi}}
   \frac{{\vert}p{\vert}}{m_{B_{c}}^{2}}
  {\vert}{\cal A}(B_{c}{\to}X_{c\bar{c}}{\pi}){\vert}^{2}
   \label{eq:br-01}
 \end{equation}
 where $X_{c\bar{c}}$ denotes $J/{\psi}$ (or ${\eta}_{c}$) for $B_{c}$ ${\to}$
 $J/{\psi}{\pi}$ (or ${\eta}_{c}{\pi}$) decays, and the momentum
 ${\vert}p{\vert}$ $=$ $(m_{B_{c}}^{2}-m_{X_{c\bar{c}}}^{2})/(2m_{B_{c}})$.
 The lifetime and mass for $B_{c}$ meson are \cite{pdg2006}:
 $m_{B_{c}}$ $=$ $6.286$ ${\pm}$ $0.005$ GeV, and
 ${\tau}_{B_{c}}$ $=$ $0.46^{+0.18}_{-0.16}$ ps.

 To perform numerical calculations, we specify the input parameters as follows.

 Nonperturbative hadronic quantities, such as meson decay constants, transition
 form factors, and meson distribution amplitudes, appear as inputs.
 In principle, information about these hadronic
 quantities can be obtained from experiments and/or estimated theoretically
 by non-perturbative method, such as lattice calculations, QCD sum rules, etc.
 In this paper, we shall follow the notation of \cite{bsw} on hadron transition form
 factors. As a good approximation in the heavy quark limit, we shall take their
 values at the maximal recoil point for discussion. From the values collected in
 Table.\ref{tab2}, we can see that there exists large uncertainty on theoretical
 predictions for $F_{0}^{B_{c}{\to}{\eta}_{c}}$ and $A_{0}^{B_{c}{\to}J/{\psi}}$.
 To estimate the branching ratios for $B_{c}$ ${\to}$ $J/{\psi}{\pi}$,
 ${\eta}_{c}{\pi}$ decays, we would like to take $F_{0}^{B_{c}{\to}{\eta}_{c}}$
 $=$ $A_{0}^{B_{c}{\to}J/{\psi}}$ $=$ $0.6$. We shall use the asymptotic form
 of the pion in the calculation, i.e. ${\phi}_{\pi}(z)$ $=$ $6z\bar{z}$.

 The Wolfenstein parameterization for the CKM matrix elements is used due to
 its advantage in explicitly expressing the hierarchy among the CKM elements
 in terms of powers of small parameter ${\lambda}$. With this notation, up
 to the order of ${\cal O}({\lambda}^{4})$, the CKM elements involved are
 $V_{cb}$ $=$ $A{\lambda}^{2}$, and $V_{ud}$ $=$ $1$ $-$ ${\lambda}^{2}/2$.
 The independent parameters of $A$ and ${\lambda}$ have been well determined
 to the accuracy of $1\%$ and $0.1\%$, respectively. Their values are \cite{pdg2006}
 \begin{equation}
  A\;=\;0.818^{+0.007}_{-0.017},~~~~~~
  {\lambda}\;=\;0.2272{\pm}0.0001
 \label{eq:ckm-elements}
 \end{equation}
 Note that the relative weak phase difference is zero at this approximation,
 so no $CP$-violating asymmetry is expected.

 Other input parameters are
 \begin{eqnarray}
  && m_{u}=m_{d}=m_{s}=0,~~~~~~~~~~~~~~~~~~~~~~~
     m_{c}=1.25{\pm}0.09~\hbox{GeV \cite{pdg2006}}
  \nonumber \\
  && m_{b}=4.20{\pm}0.07~\hbox{GeV \cite{pdg2006}}~~~~~~~~~~~~~~~~
     f_{\pi}=130.7{\pm}0.1{\pm}0.36~\hbox{MeV \cite{pdg2006}}
   \nonumber \\
  && m_{_{J/{\psi}}}=3096.916{\pm}0.011~\hbox{MeV \cite{pdg2006}}~~~~~
     m_{_{{\eta}_{c}}}=2980.4{\pm}1.2~\hbox{MeV \cite{pdg2006}}
   \nonumber
 \end{eqnarray}
 If not specified explicitly, we shall take their central values as
 the default input.

 Our numerical results are listed in Table \ref{tab1} and \ref{tab3}.

 From the numbers in Table \ref{tab1}, we can see that part information of strong
 phases is obtained by considering gluon radiative corrections to vertexes.
 Compared with the real part, the imaginary part of $a_{1}$ is small for $B_{c}$ ${\to}$
 $J/{\psi}{\pi}$, ${\eta}_{c}{\pi}$. The reason is that the ``nonfactorizable''
 effects are ${\alpha}_{s}$ or  ${\Lambda}_{QCD}/m_{b}$ suppressed
 within QCDF. Although the imaginary part is small, it will be very important for
 the exploration of $CP$ violation for decay modes receiving contributions from
 both tree and penguin topologies. In fact, the numerical results of \cite{0104110,0108141,0102077}
 indicate that the imaginary parts of most $a_{i}$ ($i$ $=$ $3,{\cdots},10$) are
 the same order as the real parts, i.e. the ``nonfactorizable'' effects cannot be
 neglected for penguin-dominated decay modes.

 It should be noted that both the imaginary part arising at the order of ${\alpha}_{s}$
 and the real part of $a_{1}$ depend on the renormalization scale. But the scale
 dependence of $a_{1}$ has been reduced compared with that of NF, at least at the order
 of ${\alpha}_{s}$. This can be seen from the Eq.(\ref{eq:scale-dependence}) and
 Fig.\ref{fig2}. In principle, high order of ${\alpha}_{s}$ corrections could be
 calculated order by order. Perhaps the remaining scale dependence of $a_{1}$ could
 be cancelled by the contributions from higher order of ${\alpha}_{s}$ corrections.

 The numerical results of the branching ratios for $B_{c}$ ${\to}$
 $J/{\psi}{\pi}$, ${\eta}_{c}{\pi}$ decays are given in Table \ref{tab3}.
 From the numbers in Table \ref{tab3}, we can see that these two branching
 ratios are close to each other, if we assume $A_{0}^{B_{c}{\to}J/{\psi}}$
 $=$ $F_{0}^{B_{c}{\to}{\eta}_{c}}$. In fact, this point can also be seen
 from their relationship
 \begin{equation}
  \frac{{\cal BR}(B_{c}{\to}J/{\psi}{\pi})}
       {{\cal BR}(B_{c}{\to}{\eta}_{c}{\pi})}{\simeq}
  \frac{{\vert}A_{0}^{B_{c}{\to}J/{\psi}}{\vert}^{2}  (m_{B_{c}}^2-m_{J/{\psi}}^2)^{3}}
       {{\vert}F_{0}^{B_{c}{\to}{\eta}_{c}}{\vert}^{2}(m_{B_{c}}^2-m_{{\eta}_{c}}^2)^{3}}
 \label{eq:br-cc}
 \end{equation}
 Some nonperturbative effects could be cancelled in the ratio
 of branching fraction. Using experimental data, the relation between the form factor
 $A_{0}^{B_{c}{\to}J/{\psi}}$ and $F_{0}^{B_{c}{\to}{\eta}_{c}}$ is expected to
 be obtained from the ratio Eq.(\ref{eq:br-cc}). Once the nonperturbative parameters
 $A_{0}^{B_{c}{\to}J/{\psi}}$ and $F_{0}^{B_{c}{\to}{\eta}_{c}}$ are fixed, we might
 get some information on CKM element $V_{cb}$ from Eq.(\ref{eq:amp-jpsi}) and
 Eq.(\ref{eq:amp-etac}).

\begin{figure}[ht]
 \includegraphics[255,380][355,600]{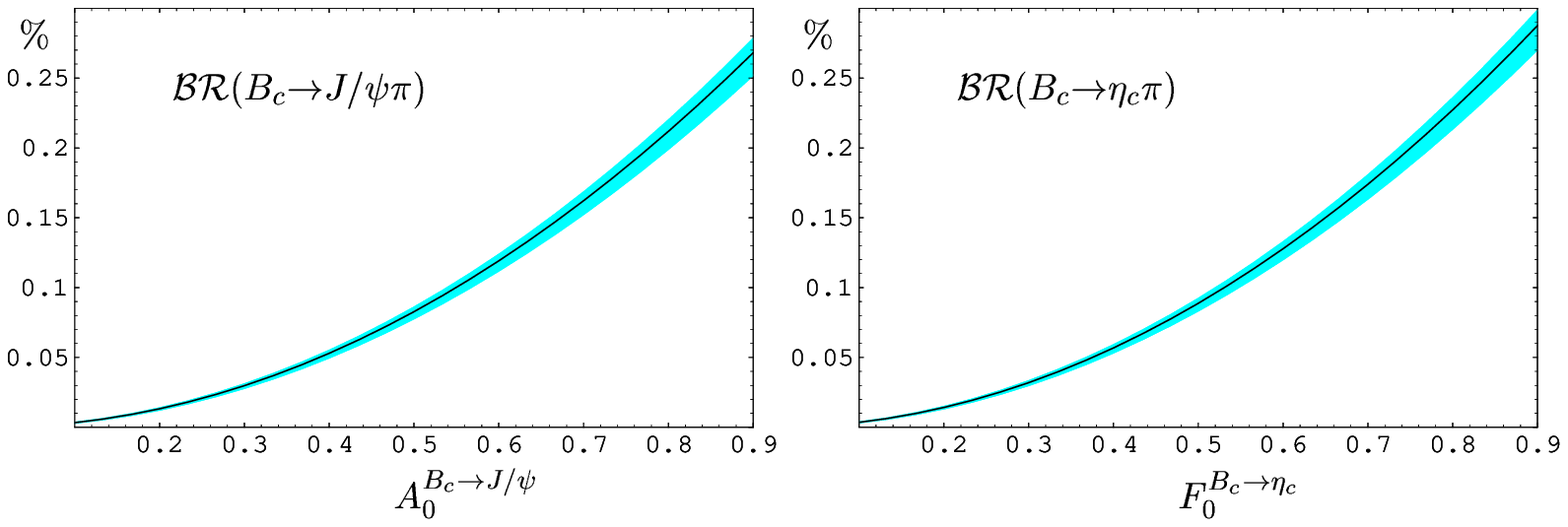}
 \caption{Branching ratios for $B_{c}$ ${\to}$ $J/{\psi}{\pi}$, ${\eta}_{c}{\pi}$
  versus the transition form factors $A_{0}^{B_{c}{\to}J/{\psi}}$,
  $F_{0}^{B_{c}{\to}{\eta}_{c}}$ in the QCDF approach, considering the mass of $c$
  quark. The solid lines are calculated with central values of inputs, while the
  bands denote the uncertainties from CKM elements and quark masses.}
 \label{fig3}
 \end{figure}

 In addition, form Table III we also see that the charm quark mass effect on the
 branching ratios is small, and that uncertainties form renormalization scale is
 also small. The uncertainties from the CKM elements which is proportional to
 $2{\sigma}_{A}/A$ are less than 5\% [see Eq.(\ref{eq:ckm-elements})].
 Thus the remaining uncertainties are the hadron parameters, such as
 transition form factors. Within QCDF approach, the transition form factor is
 the nonperturbative input parameter which comes from the long-distance contributions,
 so it must be computed nonperturbatively or determined experimentally. From the
 numbers in Table \ref{tab2}, we can see that there are large differences in the
 numerical values of transition form factors $A_{0}^{B_{c}{\to}J/{\psi}}$,
 $F_{0}^{B_{c}{\to}{\eta}_{c}}$ for various theoretical approaches.
 So at the scale of ${\mu}$ $=$ $m_{b}$, the branching ratio can be rewritten as
 \begin{eqnarray}
 &&{\cal BR}(B_{c}{\to}J/{\psi}{\pi})= 0.119\% {\times}
    \frac{{\vert}A_{0}^{B_{c}{\to}J/{\psi}}{\vert}^{2}}{0.36}
    \label{eq:br-jpsi-01} \\
 &&{\cal BR}(B_{c}{\to}{\eta}_{c}{\pi})= 0.128\% {\times}
    \frac{{\vert}F_{0}^{B_{c}{\to}{\eta}_{c}}{\vert}^{2}}{0.36}
    \label{eq:br-etac-01}
 \end{eqnarray}
 The effects of transition form factors on the branching ratios are displayed in
 Fig.\ref{fig3} \footnotemark[1], including the uncertainties from the CKM elements,
 quark masses and the renormalization scale. (Here the uncertainties from
 ${\Lambda}_{QCD}/m_{b}$ corrections is not included, but it is assumed that its contribution
 is power suppressed and should not be large \cite{0104110}) Clearly, the largest theoretical
 uncertainty comes from the transition form factors, which is of a nonperturbative nature
 within the QCD framework. Here we can not compute these form factors within QCD factorization
 scheme. We just use them as input parameters.

 \footnotetext[1]{Here, we think that theoretical prediction on nonperturbative parameters,
 such as transition form factors, relies on hard-to-quantify educated guesswork, due to our
 present inability to compute precisely strong interactions at long distances. All values within
 allowed ranges should be treated on an equal footing, irrespective of how close they are
 from the edges of the allowed range. So we had better to give a range in Fig.\ref{fig3}
 to show the theoretical uncertainties, rather than the form of Eq.(\ref{eq:ckm-elements}). }

 \section{Summary and Conclusion}
 \label{sec4}
 In this paper, we calculated the hadronic matrix element for $b$ ${\to}$ $c$
 transition at one-loop level under NDR scheme and in the heavy quark limit.
 Then we apply the master QCD formula to the would-be well detectable decay
 channels in experiments $B_{c}$ ${\to}$ $J/{\psi}{\pi}$, ${\eta}_{c}{\pi}$.
 The ``nonfactorizable'' vertex corrections are computed. We find
 that in the heavy quark limit, the ``nonfactorizable'' contributions
 dominated by hard gluon exchange can compensate the
 renormalization scale-dependence of the Wilson coefficients and are infrared
 safe and gauge independent. Finally, we give the branching ratios for
 $B_{c}$ ${\to}$ $J/{\psi}{\pi}$, ${\eta}_{c}{\pi}$ decays using the
 asymptotic distribution amplitude of the pion. We find that the large
 theoretical uncertainties come mainly from the non-perturbative transition
 form factors.

 \section*{Acknowledgments}
 This work is Supported in part both by National Natural Science Foundation
 of China under Grant No. 10647119, 10710146, 10375041, 90403024 and by Natural
 Science Foundation of Henan Province, China (Grant No. 2008B140006).
 We would like to thank Prof. Yadong Yang and Dr. Shuxian Du for valuable discussions.

 \begin{table}[ht]
 \caption{Wilson coefficients $C_{i}$ and $a_{1}$ in the NDR scheme.}
 \label{tab1}
 \begin{ruledtabular}
 \begin{tabular}{c|ll|cc|c|cc}
  ${\mu}$ & \multicolumn{2}{c|}{NLO} & \multicolumn{2}{c|}{LO} & NF & \multicolumn{2}{c}{QCDF} \\ \cline{2-8}
  (GeV) & $C_{1}$ & $C_{2}$ & $C_{1}$ & $C_{2}$ & $a_{1}$
        & \begin{tabular}{c} $a_{1}$ \\ ($m_{c}$ $=$ $1.25$ GeV) \end{tabular}
        & \begin{tabular}{c} $a_{1}$ \\ ($m_{c}$ $=$ $0$) \end{tabular} \\ \hline
 $ 2.0$ & $ 1.147$ & $-0.304$ & $ 1.191$ & $-0.374$ &
          $ 1.046$ & $ 1.084$ $+$ $i\, 0.034$ & $ 1.079$ $+$ $i\, 0.036$  \\
 $ 3.0$ & $ 1.109$ & $-0.236$ & $ 1.148$ & $-0.303$ &
          $ 1.031$ & $ 1.070$ $+$ $i\, 0.024$ & $ 1.066$ $+$ $i\, 0.025$  \\
 $ 4.0$ & $ 1.088$ & $-0.196$ & $ 1.124$ & $-0.261$ &
          $ 1.023$ & $ 1.061$ $+$ $i\, 0.019$ & $ 1.058$ $+$ $i\, 0.020$  \\
 $ 5.0$ & $ 1.074$ & $-0.169$ & $ 1.108$ & $-0.232$ &
          $ 1.018$ & $ 1.054$ $+$ $i\, 0.016$ & $ 1.052$ $+$ $i\, 0.016$  \\
 $ 6.0$ & $ 1.064$ & $-0.148$ & $ 1.096$ & $-0.210$ &
          $ 1.014$ & $ 1.049$ $+$ $i\, 0.013$ & $ 1.047$ $+$ $i\, 0.014$  \\
 $ 7.0$ & $ 1.056$ & $-0.131$ & $ 1.087$ & $-0.192$ &
          $ 1.012$ & $ 1.045$ $+$ $i\, 0.012$ & $ 1.043$ $+$ $i\, 0.012$  \\
 $ 8.0$ & $ 1.049$ & $-0.118$ & $ 1.079$ & $-0.178$ &
          $ 1.010$ & $ 1.041$ $+$ $i\, 0.011$ & $ 1.040$ $+$ $i\, 0.011$
 \end{tabular}
 \end{ruledtabular}
 \end{table}

 \begin{table}[ht]
 \caption{Values of transition form factors}
 \label{tab2}
 \begin{ruledtabular}
 \begin{tabular}{c|c|c}
 Reference         & $F_{0}^{B_{c}{\to}{\eta}_{c}}$
                   & $A_{0}^{B_{c}{\to}J/{\psi}}$ \\ \hline
 \cite{prd39p1342} & $0.170$ ${\sim}$ $0.687$ & $0.156$ ${\sim}$ $0.684$ \\
 \cite{zpc57p43}   & $0.20{\pm}0.02$ & $0.26{\pm}0.07$ \footnotemark[1] \\
 \cite{prd48p5208}   & $0.23{\pm}0.01$ & $0.21{\pm}0.03$ \footnotemark[2] \\
 \cite{npb569p473}   & $0.66$ & $0.595$ \footnotemark[3] \\
 \cite{prd63p074010} & $0.76$ & $0.69$ \footnotemark[4]  \\
 \cite{0702147}      & $0.87$ & $0.27$ \footnotemark[5] \\
 \cite{prd68p094020} & $0.47$ & $0.40$ \\
 \cite{prd71p094006} & $0.61$ & $0.57$ \footnotemark[6] \\
 \cite{07042493} &  & $0.57^{+0.01}_{-0.02}$
 \end{tabular}
 \end{ruledtabular}
 \end{table}
 \footnotetext[1]{Compared the definitions of transition form factor of
  Ref.\cite{bsw} with those of Ref.\cite{zpc57p43}, we can obtain their
  relationships at the maximal recoil point,
  \begin{equation}
  A_{0}^{B_{c}{\to}J/{\psi}}= \frac{F_{0}^{A}}{2m_{_{J/{\psi}}}}
         + \frac{m_{_{B_{c}}}^{2}-m_{_{J/{\psi}}}^{2}}{2m_{_{J/{\psi}}}}F_{+}^{A}
  \label{eq:form-relation-01}
  \end{equation}
  where the transition form factor $A_{0}^{B_{c}{\to}J/{\psi}}$ is defined in
  \cite{bsw}, $F_{0}^{A}$ and $F_{+}^{A}$ are defined in \cite{zpc57p43}, and
  their values are $F_{0}^{A}$ $=$ $2.5{\pm}0.3$ ${\rm GeV}^{-1}$,
  $F_{+}^{A}$ $=$ $-0.03{\pm}0.01$ ${\rm GeV}^{-1}$.}
 \footnotetext[2]{Using the relationship of Eq.(\ref{eq:form-relation-01})
  and the values of $F_{0}^{A}$ $=$ $2.0{\pm}0.2$ ${\rm GeV}^{-1}$,
  $F_{+}^{A}$ $=$ $-0.024{\pm}0.002$ ${\rm GeV}^{-1}$ \cite{prd48p5208}}
 \footnotetext[3]{Using the relationship of Eq.(\ref{eq:form-relation-01})
  and the values of $F_{0}^{A}$ $=$ $5.9$ ${\rm GeV}^{-1}$,
  $F_{+}^{A}$ $=$ $-0.074$ ${\rm GeV}^{-1}$ \cite{prd48p5208}}
 \footnotetext[4]{Compared the definitions of transition form factor of
  Ref.\cite{bsw} with those of Ref.\cite{prd63p074010}, we can obtain
  their relationships at the maximal recoil point,
  \begin{equation}
  A_{0}^{B_{c}{\to}J/{\psi}}=
           \frac{m_{_{B_{c}}}+m_{_{J/{\psi}}}}{2m_{_{J/{\psi}}}}A_{1}
         - \frac{m_{_{B_{c}}}-m_{_{J/{\psi}}}}{2m_{_{J/{\psi}}}}A_{2}
  \label{eq:form-relation-02}
  \end{equation}
  where $A_{1}$ and $A_{2}$ are defined in \cite{prd63p074010}, and their
  values are $A_{1}$ $=$ $0.68$, $A_{2}$ $=$ $0.66$.}
 \footnotetext[5]{Using the relationship of Eq.(\ref{eq:form-relation-02})
  and the values of $A_{1}$ $=$ $0.75$, $A_{2}$ $=$ $1.69$. \cite{0702147}}
 \footnotetext[6]{Compared the definitions of transition form factor of
  Ref.\cite{bsw} with those of Ref.\cite{prd71p094006}, we can obtain
  their relationships at the maximal recoil point,
  \begin{equation}
  A_{0}^{B_{c}{\to}J/{\psi}}=
           \frac{m_{_{B_{c}}}-m_{_{J/{\psi}}}}{2m_{_{J/{\psi}}}}
          (A_{0}-A_{+})
  \label{eq:form-relation-03}
  \end{equation}
  where $A_{0}$ and $A_{+}$ are defined in \cite{prd71p094006}, and their
  values are $A_{0}$ $=$ $1.64$, $A_{2}$ $=$ $0.54$.}

 \begin{table}[ht]
 \caption{Branching ratios for $B_{c}$ ${\to}$ $J/{\psi}{\pi}$, ${\eta}_{c}{\pi}$
  (in the unit of $\%$).}
 \label{tab3}
 \begin{ruledtabular}
 \begin{tabular}{c|ccc|ccc}
          & \multicolumn{3}{c|}{${\cal BR}(B_{c}{\to}J/{\psi}{\pi})$}
          & \multicolumn{3}{c}{${\cal BR}(B_{c}{\to}{\eta}_{c}{\pi})$} \\ \cline{2-7}
 ${\mu}$  & NF & \begin{tabular}{c} QCDF \\ ($m_{c}$ $=$ $0$) \end{tabular}
          & \begin{tabular}{c} QCDF \\ ($m_{c}$ $=$ $1.25$ GeV) \end{tabular}
          & NF & \begin{tabular}{c} QCDF \\ ($m_{c}$ $=$ $0$) \end{tabular}
          & \begin{tabular}{c} QCDF \\ ($m_{c}$ $=$ $1.25$ GeV) \end{tabular} \\ \hline
 $m_{b}/2$ & $ 0.116$ & $ 0.123$ & $ 0.124$ &
            $ 0.124$ & $ 0.132$ & $ 0.134$ \\
 $ m_{b}$ & $ 0.111$ & $ 0.119$ & $ 0.119$ &
            $ 0.119$ & $ 0.127$ & $ 0.128$ \\
 $2m_{b}$ & $ 0.108$ & $ 0.114$ & $ 0.115$ &
            $ 0.116$ & $ 0.123$ & $ 0.123$
 \end{tabular}
 \end{ruledtabular}
 \end{table}

 \end{document}